\documentclass[aps,prl,twocolumn,groupedaddress,10pt]{revtex4-1}

\usepackage{amsmath}
\usepackage{amsthm}
\usepackage{amssymb}
\usepackage{amsfonts}
\usepackage{graphicx}
\usepackage[english]{babel}
\usepackage{enumerate}
\usepackage{mathrsfs}
\usepackage{eucal}
\usepackage{url}
\usepackage{tensor}
\usepackage[T1]{fontenc}
\usepackage{subfigure}

\newcommand{\mca}{\CMcal}

\newcommand{\msf}{\mathsf}

\newcommand{\mbb}{\mathbb}

\newcommand{\tit}{\textit}

\newcommand{\txt}{\text}
\newcommand{\pde}{\partial}
\newcommand{\prp}{\perp}
\newcommand{\pel}{\parallel}

\newcommand{\eref}{\eqref}
\newcommand{\lbel}{\label}

\let\oldsqrt\sqrt
\def\sqrt{\mathpalette\DHLhksqrt}
\def\DHLhksqrt#1#2{
\setbox0=\hbox{$#1\oldsqrt{#2\,}$}\dimen0=\ht0
\advance\dimen0-0.2\ht0
\setbox2=\hbox{\vrule height\ht0 depth -\dimen0}
{\box0\lower0.4pt\box2}}

\begin{document}

\title{Model for gravitational collapse in effective quantum gravity}

\author{Andreas Kreienbuehl}
\email{a.kreienb@unb.ca}

\author{Viqar Husain}
\email{vhusain@unb.ca}

\author{Sanjeev S. Seahra}
\email{sseahra@unb.ca}

\affiliation{Department of Mathematics and Statistics, University of New Brunswick, Fredericton, NB E3B 5A3, Canada}

\begin{abstract}

We describe a non-perturbative approach to studying the gravitational collapse of a scalar field in spherical symmetry with quantum gravity corrections. Quantum effects are described by a phase space function that modifies the constraints of general relativity but does not affect the closure of their algebra. The model  may be viewed as one that incorporates a class of anomaly-free quantum gravity effects. Numerical simulations of the resulting field equations show that the model reproduces known classical results for sufficiently massive initial data but gives a dramatic change at the threshold of black hole formation characterized by oscillations and a mass gap. 

\end{abstract}

\maketitle

General relativity (GR) is incomplete as a theory of gravity. This is because the gravitational collapse of physical matter produces  a black hole with a singularity inside, which is a point in spacetime where the curvature and energy density diverge. The hope and expectation is that this problem will be overcome in a quantum theory of gravity.  Evidence for this comes  from both physical arguments and the quantization of cosmological models with singularities. 

Gravitational collapse in classical GR is well studied numerically in the spherically symmetric case
with a variety of matter fields. The first such model  was with a massless minimally coupled scalar field \cite{choptuik:93}. This pioneering work provided a mass scaling law at the onset of black hole formation which indicated that black holes form with infinitesimal mass.  It also provided a finely tuned violation of the cosmic censorship  conjecture, and a discrete self-similarity of the field variables that extends to arbitrary scales. 

These results are not expected to hold in a quantum theory of gravity because  of  quantum fluctuations of the metric in regions of sufficiently large energy density. The general understanding derived from homogeneous isotropic models is that singularity avoidance is associated with violation of energy
conditions, which in turn suggests that quantum gravity provides  a repulsive force at very short distances.  

These considerations suggest that it should be of much interest to find the quantum theory of GR coupled to the massless scalar field theory in spherical symmetry. However this model is  a non-trivial two-dimensional field theory for which no quantization is known,  in either the canonical quantization program or in the string theory approaches to quantum gravity. Given this state of affairs, we introduce and study in this letter a  model that modifies this sector of GR by taking into account a possible class of quantum effects. 

The class of modifications we consider  is motivated by the effective constraints approach  used in the loop quantum gravity (LQG)  program in the context of cosmology \cite{bojowald:06}. There are  two general types of   quantum corrections in LQG, which  come from the inverse triad and holonomy  operators  \cite{thiemann:98a}.  The work in cosmology attempted to incorporate the inverse triad corrections, and has recently been extended  to  the spherically symmetric sector in LQG  \cite{reyes:09,bojowald:09}.

In this letter we introduce a similar effective approach  in the Arnowitt-Deser-Misner (ADM)
canonical formulation.  This is motivated by the polymer quantization method applied to the spherically symmetric sector of  GR  \cite{husain:08,husain:05c}. In this formulation there is an analog of the inverse triad operator of LQG, which is the operator realization of the factor $1/\sqrt{\det([h])}$ in the Hamiltonian constraint of GR (where $h$ is  the spatial metric).   

Effective quantum gravity corrected equations based on this observation are investigated in \cite{husain:09} and \cite{ziprick:09a} where  the corrections are inserted  directly into the Einstein equations. The main result  in these works is that black holes form with a mass that is bigger than some strictly positive value.   However, in that approach it is difficult to verify   whether the implemented corrections break diffeomorphism invariance, which is a fundamental property of the theory. 

In contrast to these earlier works, the effective quantum gravity corrections studied  in this letter are via a Hamiltonian formulation, and are  such  that the Dirac constraint algebra closes. This means that the diffeomorphism  symmetry group is anomaly free. The model can be viewed as describing a deformation of the spherically symmetric sector of the GR plus scalar field theory, where the deformation comes from a class of possible quantum effects. Such an approach may be compared to the Ginzburg-Landau theory, where an effective macroscopic description of a quantum theory with some constraints is postulated.

A numerical study of the resulting field equations  reveals qualitative behaviour similar to that found in classical GR: there are sub-  and supercritical regions of initial data characterized by black hole formation and scattering, respectively. However, the scaling behaviour in the latter region is drastically different, giving a new type of oscillation and a mass gap. This  is a direct consequence of the effective quantum gravity corrections. 

We start from the ADM Hamiltonian formulation of GR with the metric given by  
\begin{equation}
	g_{\mu\nu}dx^{\mu}dx^{\nu}\equiv-N^2dt^2+A^2(N^rdt+dr)^2+B^2d\Omega^2,
\end{equation}
where $N$ is the lapse function and $N^r$ the radial and sole nonzero component of the shift vector. The functions $A$ and $B$ define the spatial line element $A^2dr^2+B^2d\Omega^2$  on the topologically fixed space $[0,\infty)\times S^2$. The symmetry reduced canonical action of this systems is
\begin{equation}
	S\equiv\int_{t_0}^{t_1}\int_0^{\infty}(p_A\dot{A}+p_B\dot{B}+p_{\phi}\dot{\phi}-\mca{H})\;dr\;dt,
\end{equation}
where the nonzero, equal-time Poisson brackets are
\begin{equation}
	\begin{gathered}
		\{A(t,r),p_A(t,\tilde{r})\}=\{B(t,r),p_B(t,\tilde{r})\}=\\
		=\{\phi(t,r),p_{\phi}(t,\tilde{r})\}=\delta(r,\tilde{r}).
	\end{gathered}
\end{equation}
The total Hamiltonian constraint density is 
\begin{equation}
	\mca{H}\equiv N\sum_{i=1}^3\mca{H}_{\prp}^{(i)}+N^r\mca{H}_{\pel r},
\end{equation}
where, having chosen the matter Lagrangian density 
\begin{equation}
	\mca{L}^{\txt{m}}\equiv-\frac{\sqrt{-|g|}}{8\pi}g^{\mu\nu}\pde_{\mu}\phi\pde_{\nu}\phi,\qquad|g|\equiv\det([g]),
\end{equation}
the Hamiltonian and diffeomorphism constraint densities respectively are
\begin{equation}\lbel{e:4}
	\begin{gathered}
		\begin{aligned}
			\mca{H}_{\perp}^{(1)}&\equiv\frac{p_A}{2B^2}(p_AA-2p_BB)\\
			&\quad-\frac{1}{2A^2}[A'B^{2\prime}-A(2BB''+B'^2)+A^3],
		\end{aligned}\\
		\mca{H}_{\perp}^{(2)}\equiv\frac{p_{\phi}^2}{2AB^2},\qquad\mca{H}_{\perp}^{(3)}\equiv\frac{B^2\phi'^2}{2A},\\
		\mca{H}_{\pel r}\equiv-p_A'A+p_BB'+p_{\phi}\phi'.
	\end{gathered}
\end{equation}
Here, the part $\mca{H}_{\prp}^{(1)}$ of the Hamiltonian constraint density provides intrinsic and extrinsic geometric information of $[0,\infty)\times S^2$, the part $\mca{H}_{\prp}^{(2)}$ describes the kinetic energy of $\phi$, and the part $\mca{H}_{\prp}^{(3)}$ its gradient energy.

We note that  $\mca{H}_{\prp}^{(1)}$ and $\mca{H}_{\prp}^{(2)}$ are proportional  to $1/B^2$,  because this factor comes from the inverse determinant of the spatial metric. The classical fall-off conditions on $B$ in the asymptotic region imply that this factor is singular at $r=0$.  However, in the corresponding quantum theory \cite{husain:08} the  operator $\widehat{1/B^2}$ has a bounded spectrum on the polymer Hilbert space.  These considerations together with those in  \cite{bojowald:06} motivate the effective constraint density
\begin{equation}\lbel{e:5}
	\mca{H}^{\txt{e}}\equiv N\sum_{i=1}^3Q_{(i)}\mca{H}_{\prp}^{(i)}+N^r\mca{H}_{\pel r},
\end{equation}
where the functions $Q_{(i)}\equiv Q_{(i)}(B)$ are designed to render the factors $1/B^2$ regular. The functions $Q_{(i)}$ can be thought of as providing effective quantum gravity corrections, using the heuristic guidance that  operators are replaced by expectation values in suitable states. 

Given the modification of the constraints it is essential to check their algebra.  If the condition
\begin{equation}
	Q_{(1)}^2=Q_{(2)}Q_{(3)}
\end{equation}
is satisfied,  the Dirac algebra, that is here the algebra generated by the modified constraint densities, closes and is anomaly-free. This implies an invariance under diffeomorphisms, that all degrees of freedom are accounted for, and that the ADM energy is conserved.

Since $\mca{H}_{\prp}^{(3)}$ is not proportional to $1/B^2$ the function $Q_{(3)}$ can be set equal to $1$. To further simplify Hamilton's equations of motion generated by $\mca{H}^{\txt{e}}$, the Schwarzschild (or polar areal) gauges $B\equiv r$ and $p_A\equiv0$ can be fixed. The line element then takes the familiar form
\begin{equation}
	g_{\mu\nu}dx^{\mu}dx^{\nu}=-N^2dt^2+A^2dr^2+r^2d\Omega^2
\end{equation}
and the field equations become
\begin{equation}\lbel{e:2}
	\begin{gathered}
		\frac{A'}{A}=\frac{1}{2r}\left(1-A^2+\frac{p_{\phi}^2+q^4\phi'^2}{q^2}\right),\\
		\frac{s'}{s}=\frac{1-A^2}{r},\\
		\dot{\phi}=\frac{p_{\phi}}{q^2s},\qquad\dot{p}_{\phi}=\left(\frac{q^2\phi'}{s}\right)'.
	\end{gathered}
\end{equation}
The function
\begin{equation}\lbel{e:6}
	q\equiv\frac{r}{\sqrt{Q_{(1)}}}
\end{equation}
encodes the remaining quantum gravity modification and $s\equiv q^2A/(r^2N)$ is the characteristic speed of the scalar field. 

 If $q=r$   the modified equations in \eref{e:2} reduce to Einstein's equations, which are studied in \cite{choptuik:93} with the result that the black hole mass $M$ scales with the amplitude $\phi_0$ of the initial scalar field profile according to 
\begin{equation}\lbel{e:3}
	M\propto|\phi_0-\phi_0^*|^{\gamma}.
\end{equation}
In this relation $\phi_0^*$ is such that $\phi_0>\phi_0^*$ gives rise to a black hole whereas $\phi_0<\phi_0^*$ does not. If $\phi_0=\phi_0^*$ a naked singularity is the result of the collapse. The critical exponent $\gamma\simeq0.374$ \cite{choptuik:93,gundlach:97} is universal in the sense that it is  insensitive to changes of the initial scalar field data.

For the modified equations, the numerical analysis is simplified in coordinates $(u,v)$ adapted to the characteristics $c_{\pm}^{\mu}\pde_{\mu}\equiv s\pde_t\pm\pde_r$ of the scalar field.  The norm of these 
characteristics is  
\begin{equation}
	  g(c_{\pm},c_{\pm})=-\left[\left(\frac{q}{r}\right)^4-1\right]A^2,
\end{equation}
so that lines of constant $u,v\in\mbb{R}$ define spacelike directions if $q<r$, null directions if $q=r$, and timelike directions if $q>r$. That is, the scalar field follows null lines of the metric if and only if the modifications are switched off. This  makes possible a violation of the dominant energy condition.  

Setting $A^2\equiv F/f$ the field equations in \eref{e:2} can be shown to be equivalent to
\begin{equation}\lbel{e:1}
	\begin{gathered}
		\pde_ur=-\frac{f}{2},\\
		\pde_u\Phi=\frac{1}{2r}\left[\left(1-\frac{rq''}{q'}\right)f-F\right](\phi-\Phi),\\
		\pde_v\phi=-\frac{\pde_vrq'}{q}(\phi-\Phi),\\
		\pde_vF=\frac{\pde_vrq'^2}{r}F(\phi-\Phi)^2,\\
		\pde_vf=-\frac{\pde_vr}{r}(f-F),
	\end{gathered}
\end{equation}
in characteristic coordinates $(u,v)$ \cite{kreienbuehl:10}. If $q=r$ these are once again Einstein's equations but this time in double null coordinates. (They have been numerically studied in \cite{garfinkle:94} with an economical method.)  The difference between the characteristic equations in \eref{e:1} and the effective quantum gravity corrected double null equations in \cite{husain:09} is that here the equations  originate from a Hamiltonian formulation. In the remainder of the letter numerical solutions to the equations in \eref{e:1} are constructed,  with the aim of outlining how the classical  mass scaling law \eref{e:3} is modified.

\begin{figure}[h]
	\includegraphics{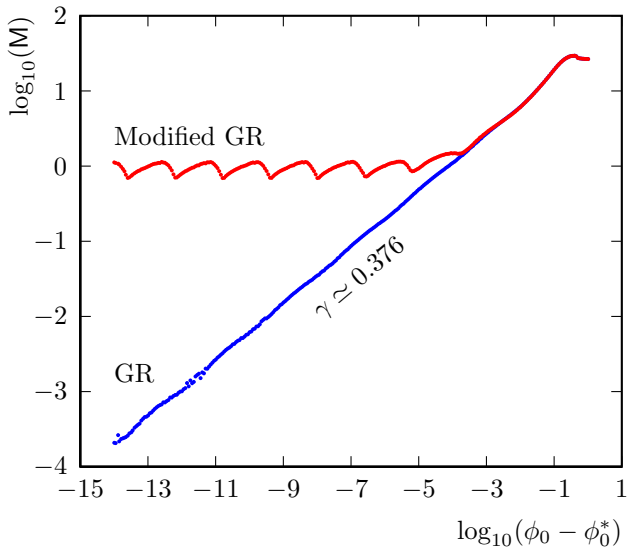}
	\caption{\lbel{f:1}An illustration of the classical and the modified mass scaling law. The relevant features are a mass gap, non-sinusoidal oscillations, an overlap with the classical data, and a scaling with $\lambda$.}
\end{figure}
\begin{figure*}
	\subfigure[\lbel{f:21}Deformation function.]{\includegraphics{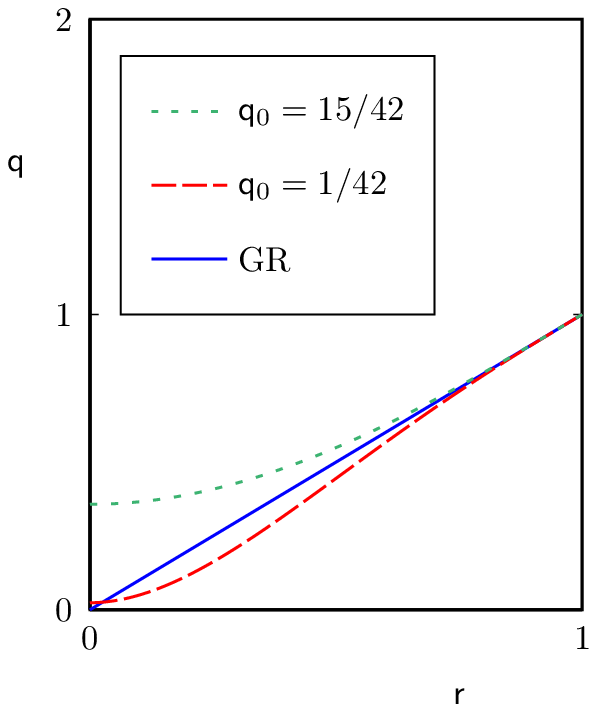}}\qquad\qquad
	\subfigure[\lbel{f:22}Mass scaling.]{\includegraphics{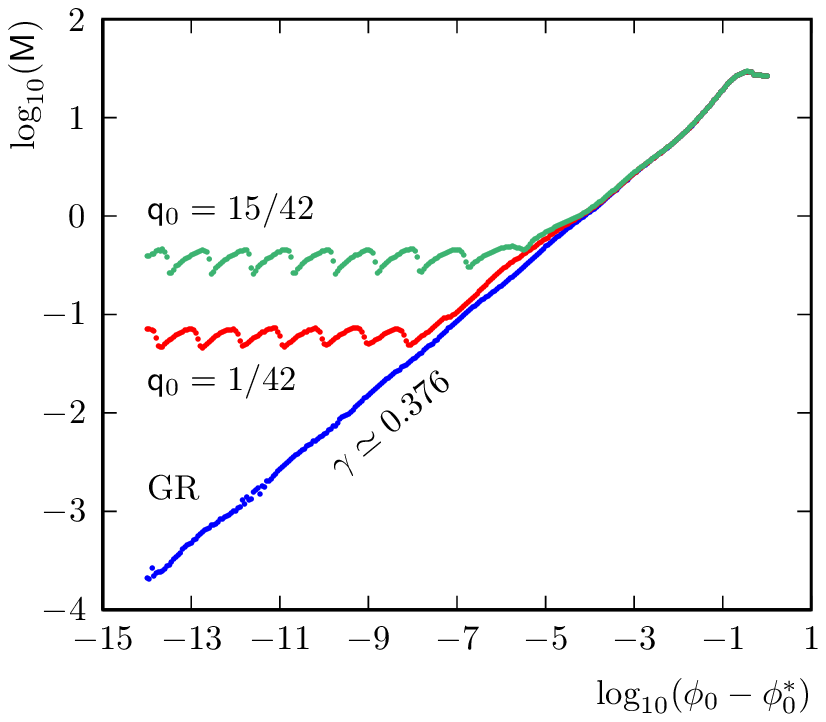}}
	\caption{\lbel{f:2}Two members of a family of functions $\msf{q}$ are displayed in Figure \ref{f:21}. For $\msf{r}<1$ the parameter $\msf{q}_0=1/4$ is at the interface between space- and timelike slices. The mass scaling data in Figure \ref{f:22} displays the effect of a change in $\msf{q}_0$ and implies a robustness of the general features.}
\end{figure*}

At this point the choice of the deformation function $q$ remains to be made. It is restricted by the demand that $Q_{(1)}/r^2$ be regular at $r=0$  and the intuition that quantum gravity corrections vanish for  $r\to\infty$. A possible choice is therefore $q\equiv\sqrt{\lambda^2+r^2}$, where $\lambda$ is a length scale that specifies the extent of the correction. This parameter may be thought of as 
related to the Planck scale which  is inherent in the  polymer quantization method.  For the following analysis it is helpful to introduce the dimensionless variables
\begin{equation}
	\msf{u}\equiv u/\lambda,\qquad\msf{v}\equiv v/\lambda,\qquad\msf{r}\equiv r/\lambda,\qquad\msf{q}\equiv q/\lambda,
\end{equation}
 which also render the field equations in \eref{e:1} dimensionless. The deformation function $q$ is now
\begin{equation}
	\msf{q}=\sqrt{1+\msf{r}^2}.
\end{equation}

The numerical method is based on a \tit{uniform grid code} that uses the unique combination of the Richtmyer two-step Lax-Wendroff method and the second-order Runge-Kutta method to solve the equations in \eref{e:1}. That is, unlike in previous work, \tit{no use of a mesh refinement is made}. The boundary conditions are $\msf{r}(\msf{u},\msf{u})=0$ and that $\Phi$ is regular at $\msf{v}=\msf{u}$. The initial data is $\msf{r}(0,\msf{v})\equiv\msf{v}/2$ for the geometric sector and
\begin{equation}\lbel{e:7}
	\phi(0,\msf{v})\equiv\phi_0\frac{\msf{v}^3}{1+\msf{v}^3}\exp\left(-\frac{(\msf{v}-\msf{v}_0)^2}{\msf{w}^2}\right)
\end{equation}
for the matter sector. 

Black hole formation is signalled when the outward null expansion $\theta\equiv\sqrt{2f/F}/\msf{r}$ approaches $0$. The last recorded minimal value of $\theta$ defines a point $(\msf{u},\msf{v})$ on the computational domain, which in turn defines the mass $\msf{M}\equiv\msf{r}(\msf{u},\msf{v})/2$. Figure \ref{f:1} displays the results of the analysis. It shows both the classical GR and  quantum modified mass scaling law \eref{e:3}. The main features of the latter are (i) a strictly positive lower bound on the mass $\msf{M}$,  (ii) oscillations, (iii) an overlap with the classical data for $\log_{10}(\phi_0-\phi_0^*)\gtrsim-4$, and (iv) a scaling with the   length scale $\lambda$ (recall that $\msf{M}\equiv M/\lambda$).

These  features  are ``universal''  in the sense that they are unaltered if the initial scalar field profile \eref{e:7} is replaced by
\begin{equation}
	\phi(0,\msf{v})\equiv\phi_0\tanh\left(\frac{\msf{v}-\msf{v}_0}{\msf{w}}\right).
\end{equation}
The value of the mass gap as well as the period and the amplitude of the oscillations agrees with the corresponding value in Figure \ref{f:1} up to a relative error of about $1.5\%$ \cite{kreienbuehl:11a}.  

Another aspect of the universality is the apparent independence of the results on the deformation function $\msf{q}$. To see this, a one-parameter family of twice differentiable functions $\msf{q}$ can be defined by
\begin{equation}\lbel{e:8}
	\msf{q}\equiv\begin{cases}
		\begin{aligned}
			&\msf{q}_0+3(1-2\msf{q}_0)\msf{r}^2\\
			&+(-3+8\msf{q}_0)\msf{r}^3+(1-3\msf{q}_0)\msf{r}^4,
		\end{aligned}
		&\msf{r}<1,\\
		\msf{r},&\msf{r}\geq1.
	\end{cases}
\end{equation}
For clarity it is chosen to reduce to the case of GR for $\msf{r}\geq1$ and it is plotted in Figure \ref{f:21} for two sample values of $\msf{q}_0\equiv\msf{q}(0)$ (which must be less than or equal to $1/2$ for regularity of the equations in \eref{e:1}). The mass scaling relations displayed in Figure \ref{f:22} exhibit the same features as in Figure \ref{f:1}. This suggests that the results presented here are not only universal with respect to different initial data profiles but also \tit{forminvariant under changes of the deformation function}.

In summary, the main developments reported here are (i) an application of the effective constraints
method \cite{bojowald:06} to spherically symmetric  gravitational collapse in the ADM formalism motivated by the polymer quantization of this sector \cite{husain:08}, and (ii) a numerical simulation of the resulting equations. 

The results indicate that the process of black hole formation is significantly altered if quantum gravity effects are taken into account: black holes form with a non-zero mass and this mass stays 
constant up to the observed oscillations and until the collapsing matter reaches a certain threshold. Beyond this  point the classical results are recovered.  It is interesting that the requirement of regularity of the inverse metric function factors, combined with closure of the constraint algebra, lead to a universality of the results with respect to both the initial data and  the deformation function. 

It would be interesting to perform a  careful analysis of the modified critical solution. Classically, 
this solution is a naked singularity.  In the effective theory it is possible that this is replaced by an
unstable ``boson star'' solution  which sits at the threshold of the supercritical regime.  

\smallskip
\begin{acknowledgments}
This work was supported in part by the Natural Science and Engineering Research Council of Canada. We thank Martin Bojowald, Juan Reyes, Jack Gegenberg, Gabor Kunstatter, and David Sloan for discussions, and the Atlantic Computational Excellence Network for computing resources.
\end{acknowledgments}


\bibliographystyle{apsrev4-1}
\bibliography{/home/akreienb/latex/biblio/b}

\begin{thebibliography}{13}%
\makeatletter
\providecommand \@ifxundefined [1]{%
 \@ifx{#1\undefined}
}%
\providecommand \@ifnum [1]{%
 \ifnum #1\expandafter \@firstoftwo
 \else \expandafter \@secondoftwo
 \fi
}%
\providecommand \@ifx [1]{%
 \ifx #1\expandafter \@firstoftwo
 \else \expandafter \@secondoftwo
 \fi
}%
\providecommand \natexlab [1]{#1}%
\providecommand \enquote  [1]{``#1''}%
\providecommand \bibnamefont  [1]{#1}%
\providecommand \bibfnamefont [1]{#1}%
\providecommand \citenamefont [1]{#1}%
\providecommand \href@noop [0]{\@secondoftwo}%
\providecommand \href [0]{\begingroup \@sanitize@url \@href}%
\providecommand \@href[1]{\@@startlink{#1}\@@href}%
\providecommand \@@href[1]{\endgroup#1\@@endlink}%
\providecommand \@sanitize@url [0]{\catcode `\\12\catcode `\$12\catcode
  `\&12\catcode `\#12\catcode `\^12\catcode `\_12\catcode `\%12\relax}%
\providecommand \@@startlink[1]{}%
\providecommand \@@endlink[0]{}%
\providecommand \url  [0]{\begingroup\@sanitize@url \@url }%
\providecommand \@url [1]{\endgroup\@href {#1}{\urlprefix }}%
\providecommand \urlprefix  [0]{URL }%
\providecommand \Eprint [0]{\href }%
\providecommand \doibase [0]{http://dx.doi.org/}%
\providecommand \selectlanguage [0]{\@gobble}%
\providecommand \bibinfo  [0]{\@secondoftwo}%
\providecommand \bibfield  [0]{\@secondoftwo}%
\providecommand \translation [1]{[#1]}%
\providecommand \BibitemOpen [0]{}%
\providecommand \bibitemStop [0]{}%
\providecommand \bibitemNoStop [0]{.\EOS\space}%
\providecommand \EOS [0]{\spacefactor3000\relax}%
\providecommand \BibitemShut  [1]{\csname bibitem#1\endcsname}%
\let\auto@bib@innerbib\@empty
\bibitem [{\citenamefont {Choptuik}(1993)}]{choptuik:93}%
  \BibitemOpen
  \bibfield  {author} {\bibinfo {author} {\bibfnamefont {M.~W.}\ \bibnamefont
  {Choptuik}},\ }\href@noop {} {\bibfield  {journal} {\bibinfo  {journal}
  {Phys. Rev. Lett.}\ }\textbf {\bibinfo {volume} {70}},\ \bibinfo {pages} {9}
  (\bibinfo {year} {1993})}\BibitemShut {NoStop}%
\bibitem [{\citenamefont {Bojowald}\ \emph {et~al.}(2006)\citenamefont
  {Bojowald}, \citenamefont {Kagan}, \citenamefont {Singh}, \citenamefont
  {Hernandez},\ and\ \citenamefont {Skirzewski}}]{bojowald:06}%
  \BibitemOpen
  \bibfield  {author} {\bibinfo {author} {\bibfnamefont {M.}~\bibnamefont
  {Bojowald}}, \bibinfo {author} {\bibfnamefont {M.}~\bibnamefont {Kagan}},
  \bibinfo {author} {\bibfnamefont {P.}~\bibnamefont {Singh}}, \bibinfo
  {author} {\bibfnamefont {H.~H.}\ \bibnamefont {Hernandez}}, \ and\ \bibinfo
  {author} {\bibfnamefont {A.}~\bibnamefont {Skirzewski}},\ }\href@noop {}
  {\bibfield  {journal} {\bibinfo  {journal} {Phys. Rev. D}\ }\textbf {\bibinfo
  {volume} {74}},\ \bibinfo {pages} {123512} (\bibinfo {year}
  {2006})}\BibitemShut {NoStop}%
\bibitem [{\citenamefont {Thiemann}(1998)}]{thiemann:98a}%
  \BibitemOpen
  \bibfield  {author} {\bibinfo {author} {\bibfnamefont {T.}~\bibnamefont
  {Thiemann}},\ }\href@noop {} {\bibfield  {journal} {\bibinfo  {journal}
  {Class. Quantum Grav.}\ }\textbf {\bibinfo {volume} {15}},\ \bibinfo {pages}
  {839} (\bibinfo {year} {1998})}\BibitemShut {NoStop}%
\bibitem [{\citenamefont {Reyes}(2009)}]{reyes:09}%
  \BibitemOpen
  \bibfield  {author} {\bibinfo {author} {\bibfnamefont {J.~D.}\ \bibnamefont
  {Reyes}},\ }\emph {\bibinfo {title} {{Spherically Symmetric Loop Quantum
  Gravity: Connection to Two-Dimensional Models and Applications to
  Gravitational Collapse}}},\ \href@noop {} {Ph.D. thesis},\ \bibinfo  {school}
  {The Pennsylvania State University} (\bibinfo {year} {2009})\BibitemShut
  {NoStop}%
\bibitem [{\citenamefont {Bojowald}\ \emph {et~al.}(2009)\citenamefont
  {Bojowald}, \citenamefont {Reyes},\ and\ \citenamefont
  {Tibrewala}}]{bojowald:09}%
  \BibitemOpen
  \bibfield  {author} {\bibinfo {author} {\bibfnamefont {M.}~\bibnamefont
  {Bojowald}}, \bibinfo {author} {\bibfnamefont {J.~D.}\ \bibnamefont {Reyes}},
  \ and\ \bibinfo {author} {\bibfnamefont {R.}~\bibnamefont {Tibrewala}},\
  }\href@noop {} {\bibfield  {journal} {\bibinfo  {journal} {Phys. Rev. D}\
  }\textbf {\bibinfo {volume} {80}},\ \bibinfo {pages} {084002} (\bibinfo
  {year} {2009})}\BibitemShut {NoStop}%
\bibitem [{\citenamefont {Husain}\ and\ \citenamefont
  {Winkler}(2005{\natexlab{a}})}]{husain:08}%
  \BibitemOpen
  \bibfield  {author} {\bibinfo {author} {\bibfnamefont {V.}~\bibnamefont
  {Husain}}\ and\ \bibinfo {author} {\bibfnamefont {O.}~\bibnamefont
  {Winkler}},\ }\href@noop {} {\bibfield  {journal} {\bibinfo  {journal}
  {Class. Quantum Grav.}\ }\textbf {\bibinfo {volume} {22}},\ \bibinfo {pages}
  {L127} (\bibinfo {year} {2005}{\natexlab{a}})}\BibitemShut {NoStop}%
\bibitem [{\citenamefont {Husain}\ and\ \citenamefont
  {Winkler}(2005{\natexlab{b}})}]{husain:05c}%
  \BibitemOpen
  \bibfield  {author} {\bibinfo {author} {\bibfnamefont {V.}~\bibnamefont
  {Husain}}\ and\ \bibinfo {author} {\bibfnamefont {O.}~\bibnamefont
  {Winkler}},\ }\href@noop {} {\bibfield  {journal} {\bibinfo  {journal}
  {Class. Quantum Grav.}\ }\textbf {\bibinfo {volume} {22}},\ \bibinfo {pages}
  {L135} (\bibinfo {year} {2005}{\natexlab{b}})}\BibitemShut {NoStop}%
\bibitem [{\citenamefont {Husain}(2009)}]{husain:09}%
  \BibitemOpen
  \bibfield  {author} {\bibinfo {author} {\bibfnamefont {V.}~\bibnamefont
  {Husain}},\ }\href@noop {} {\bibfield  {journal} {\bibinfo  {journal} {Adv.
  Sci. Lett.}\ }\textbf {\bibinfo {volume} {2}},\ \bibinfo {pages} {214}
  (\bibinfo {year} {2009})}\BibitemShut {NoStop}%
\bibitem [{\citenamefont {Ziprick}\ and\ \citenamefont
  {Kunstatter}(2009)}]{ziprick:09a}%
  \BibitemOpen
  \bibfield  {author} {\bibinfo {author} {\bibfnamefont {J.}~\bibnamefont
  {Ziprick}}\ and\ \bibinfo {author} {\bibfnamefont {G.}~\bibnamefont
  {Kunstatter}},\ }\href@noop {} {\bibfield  {journal} {\bibinfo  {journal}
  {Phys. Rev. D}\ }\textbf {\bibinfo {volume} {80}},\ \bibinfo {pages} {024032}
  (\bibinfo {year} {2009})}\BibitemShut {NoStop}%
\bibitem [{\citenamefont {Gundlach}(1997)}]{gundlach:97}%
  \BibitemOpen
  \bibfield  {author} {\bibinfo {author} {\bibfnamefont {C.}~\bibnamefont
  {Gundlach}},\ }\href@noop {} {\bibfield  {journal} {\bibinfo  {journal}
  {Phys. Rev. D}\ }\textbf {\bibinfo {volume} {55}},\ \bibinfo {pages} {695}
  (\bibinfo {year} {1997})}\BibitemShut {NoStop}%
\bibitem [{\citenamefont {Kreienbuehl}\ \emph {et~al.}(2010)\citenamefont
  {Kreienbuehl}, \citenamefont {Husain},\ and\ \citenamefont
  {Seahra}}]{kreienbuehl:10}%
  \BibitemOpen
  \bibfield  {author} {\bibinfo {author} {\bibfnamefont {A.}~\bibnamefont
  {Kreienbuehl}}, \bibinfo {author} {\bibfnamefont {V.}~\bibnamefont {Husain}},
  \ and\ \bibinfo {author} {\bibfnamefont {S.~S.}\ \bibnamefont {Seahra}},\
  }\href@noop {} {\enquote {\bibinfo {title} {{Modified general relativity as a
  model for quantum gravitational collapse}},}\ } (\bibinfo {year} {2010}),\
  \bibinfo {note} {arXiv:1011.2381v1}\BibitemShut {NoStop}%
\bibitem [{\citenamefont {Garfinkle}(1994)}]{garfinkle:94}%
  \BibitemOpen
  \bibfield  {author} {\bibinfo {author} {\bibfnamefont {D.}~\bibnamefont
  {Garfinkle}},\ }\href@noop {} {\bibfield  {journal} {\bibinfo  {journal}
  {Phys. Rev. D}\ }\textbf {\bibinfo {volume} {51}},\ \bibinfo {pages} {5558}
  (\bibinfo {year} {1994})}\BibitemShut {NoStop}%
\bibitem [{\citenamefont {Kreienbuehl}\ \emph {et~al.}(2011)\citenamefont
  {Kreienbuehl}, \citenamefont {Husain},\ and\ \citenamefont
  {Seahra}}]{kreienbuehl:11a}%
  \BibitemOpen
  \bibfield  {author} {\bibinfo {author} {\bibfnamefont {A.}~\bibnamefont
  {Kreienbuehl}}, \bibinfo {author} {\bibfnamefont {V.}~\bibnamefont {Husain}},
  \ and\ \bibinfo {author} {\bibfnamefont {S.~S.}\ \bibnamefont {Seahra}},\
  }\href@noop {} {} (\bibinfo {year} {2011}),\ \bibinfo {note} {in
  preparation}\BibitemShut {NoStop}%
\end{thebibliography}%

\end{document}